\documentstyle[12pt,twoside,fleqn,espcrc1,epsfig]{article}

\newcommand{\AmS}{{\protect\the\textfont2
  A\kern-.1667em\lower.5ex\hbox{M}\kern-.125emS}}

\title{Vector mesons in nuclear medium with small three momentum, a
                                     QCD sum rule approach }

\author{Su Houng Lee\address{GSI, Planckstr. 1, D-64291 Darmstadt,
 Germany}\address{Department of Physics, Yonsei University, 
Seoul 120-749,    Korea}\thanks{AvH Fellow.}
  and Hungchong Kim\address{Department of Physics, Tokyo Institute of 
Technology, Oh-Okayama, Meguro, Tokyo 152, Japan}
}
\begin{document}
% typeset front matter
\maketitle

\begin{abstract}
Using the QCD Operator Product Expansion, we  derive the real part  
of the transverse and longitudinal 
 vector vector correlation function with the $\rho,\omega$
quantum numbers to leading order in density and in ${\bf q}^2$ at $-\omega^2
\rightarrow \infty $.  To dimension 6, only twist-2 and 4 operators contribute. 
These OPE, through the  energy  dispersion relation, 
provide  model independent constraints  for the ${\bf q}$ dependence of the 
vector meson spectral density in nuclear  medium.
 We further make a  QCD sum rule type of 
analysis to extract the momentum dependence of the vector meson dispersion
relation in medium.  The contributions from twist-2 operators are added up 
to infinite order to check the validity of the OPE at the relevant Borel window.  
\end{abstract}

\section{Introduction}

The properties of vector meson in nuclear medium have been the focus of 
current interest due to its potential role to provide one with a direct 
observable of the  nuclear medium effects, associated with chiral 
symmetry restoration, via dileptons in H-A or A-A reactions\cite{L98}.   
Indeed dileptons from Relativistic Heavy 
Ion Collisions (RHIC)\cite{CERES} seemed to suggest a non-trivial change of 
the 
vector meson spectral density in a hot/dense environment, which can be 
understood in terms of model calculations\cite{LKB95} based on 
decreasing vector meson masses in hot/dense medium\cite{BR91,HL92}.  
However, model calculations\cite{RCW97,KW98} based on changes of the 
vector meson spectral densities using hadronic variables also seem to 
explain the main features of the CERES data.  
In all of the approaches, the central question is, how the
spectral density changes in hot/dense matter\cite{BR98}.  
In this talk, I will provide constraints on the three momentum (${\bf q}$)
 dependence of the vector meson spectral density in nuclear  
medium\cite{L98c,LM98}.
The approach is based on the Operator Product Expansion (OPE) in QCD and 
provides  model independent constraints that can be used to check 
the validity of any model calculation.
In the second stage, I will make a simple ansatz for the spectral density 
near the vector meson mass 
and look at what the  parameters of the ansatz should be to satisfy the 
constraints\cite{L98c}.

\section{OPE}

Consider the correlation function of the  vector current 
 $J_\mu=\bar{q} \gamma_\mu q$ in nuclear matter;
\begin{eqnarray}
\Pi_{\mu\nu} (\omega, {\bf q} ) &=& i \int d^4x e^{iqx}\langle G|
      T [ J_\mu(x) J_\nu(0) ] |G \rangle.
\label{ope1}
\end{eqnarray}
Here $ |G \rangle$ is the nuclear ground state at rest and  
 $q=(\omega,{\bf q})$.
 In what follows, when we give result for explicit vector meson, we will
use the currents  $J_\mu^{\rho,\omega}=\frac{1}{2} ( {\bar u} \gamma_\mu 
 u \mp {\bar d} \gamma_\mu d )$ for the $\rho,\omega$ mesons.

In general, because the vector current is conserved, 
 the polarization tenser in  eq.(\ref{ope1}) will have only two invariant
 functions\cite{tensors}. 
\begin{eqnarray}
\Pi_{\mu\nu}(\omega,{\bf q})=\Pi_T q^2 {\rm P}^T_{\mu\nu}+ \Pi_L q^2 
 {\rm P}^L_{\mu\nu},
\label{ope2}
\end{eqnarray}   
where we assume the ground state to be at rest, such that, $
{\rm P}^T_{00} =  {\rm P}^T_{0i}={\rm P}^T_{i0}=0$,  
 ${\rm P}^T_{ij}  =  \delta_{ij}-{\bf q}_i {\bf q}_j/{\bf q}^2 $ and $
{\rm P}^L_{\mu\nu}  =  (q_\mu q_\nu/q^2-g_{\mu\nu}- {\rm P}^T_{\mu\nu})$. 
When ${\bf q} \rightarrow 0$, 
$\Pi_L=\Pi_T$, as in the vacuum.

In this work, we are only interested in the small three momentum 
 (${\bf q}$) dependence of the 
polarization functions.   Therefore we will 
 make a small  ${\bf q}$ expansion of the correlation function and  look 
at its  energy dispersion relation at fixed ${\bf q}$,
\begin{eqnarray}
{\rm Re} \Pi_{L,T}(\omega^2,{\bf q}^2)  =  {\rm Re} \left( \Pi^0(\omega^2,0)+ 
\Pi_{L,T}^1(\omega^2,0) ~ {\bf q}^2 + \cdot \cdot \right) \nonumber \\  
  =  \int_0^\infty du^2 \left(
 {\rho(u,0)_{L,T}^0 \over (u^2-\omega^2)} +  {\rho(u,0)_{L,T}^1 \over (u^2-\omega^2)} ~ {\bf q}^2 + \cdot \cdot \right),
\label{ope3}
\end{eqnarray}
where $\rho(u,{\bf q})=1/\pi {\rm Im} \Pi^R(u^2 ,{\bf q}^2)$,  and 
$R$ denotes the retarded correlation function.  
 The OPE for $\Pi^0$ and its change in nuclear medium together with its 
corresponding energy dispersion relation provide constraints on the 
mass shift and spectral changes at ${\bf q}=0$\cite{HL92}.  
In this talk, we will discuss the 
three momentum dependence by studying the OPE and its changes in medium for 
 $\Pi_{L,T}^1$.

In general, the OPE \cite{Wilson69,muta} 
 for the polarization function at $Q^2=-\omega^2+{\bf q}^2 \rightarrow large$
will look as follows.

\begin{eqnarray}
\label{ope4}
\Pi_{\mu \nu}(\omega,{\bf q}) &  = & 
(q_\mu q_\nu-g_{\mu \nu} q^2) 
\left[ -c_0 {\rm ln}|Q^2|+ \sum_n {1 \over Q^n} A^{n,n} \right] 
\nonumber \\[12pt]
& & + \sum_{\tau=2} \sum_{k=1} [-g_{\mu \nu} q_{\mu_1} q_{\mu_2}
+g_{\mu \mu_1} q_\nu q_{\mu_2}+q_\mu q_{\mu_1}g_{\nu \mu_2} +
g_{\mu \mu_1} g_{\nu \mu_2} Q^2] \nonumber \\
& & ~~~~~~~ q_{\mu_3} \cdot \cdot  q_{\mu_{2k}}
\frac{2^{2k}}{Q^{4k+\tau-2}} A^{2k+\tau,\tau}_{\mu_1 \cdot \cdot \mu_{2k}}
\nonumber \\[12pt]
& & + \sum_{\tau=2} \sum_{k=1} [g_{\mu \nu}- q_\mu q_\nu/q^2]
q_{\mu_3} \cdot \cdot  q_{\mu_{2k}}
\frac{2^{2k}}{Q^{4k+\tau-2}} C^{2k+\tau,\tau}_{\mu_1 \cdot \cdot \mu_{2k}}.
\end{eqnarray}

Here we have pulled out the trivial $\frac{q^\alpha \cdot \cdot}{ Q^n}$ 
dependence so that $A^{d,\tau},C^{d,\tau}$ represents the residual Wilson 
coefficient 
times  matrix element of an operator of 
 dimension $d$ and twist $\tau=d-s$, where $s=2k$ is the number of spin 
index of the operator.  The first set of terms comes from the OPE of 
scalar operators, the second set from  operators with spin, which have been 
written as a double sum of twist and spin.    
 $A,C$ represents the two linearly independent sum of operators, which 
 reflect the two linearly independent polarization  
directions.

The longitudinal and transverse polarization functions can be obtained from  
 eq.(\ref{ope4}).   
The ${\bf q}$ dependence coming from the first line of  eq.(\ref{ope4}),
namely the contribution from the scalar operators,  comes from the  ${\bf q}$
dependence in  $Q^2$.
This forms the so called ``trivial'' ${\bf q}$ dependence, 
and comes from  replacing $\omega^2 \rightarrow \omega^2-{\bf q}^2$ when 
going from zero to finite three momentum.  
Here, we are not interested in these trivial dependence  and will only 
look at the ``non-trivial'' ${\bf q}$ dependence in $\Pi_{L,T}^1$.  
  Consequently  the scalar operators do not contribute to $\Pi_{L,T}^1$. 
Only operators with spins contribute to the 
 non-trivial ${\bf q}$ dependence.   
 A  prescription to find the nontrivial ${\bf q}^2$ 
dependence in the  OPE is to first calculate the total contribution 
proportional to ${\bf q}^2$ and  
then subtract out the trivial dependence; $
\frac{1}{\omega^n}(d) \frac{{\bf q}^2}{\omega^2}
~~\rightarrow (d-\frac{n}{2})
\frac{{\bf q}^2}{\omega^{n+2}}$, 
and its contribution to $\Pi^1$ would be $
(d-\frac{n}{2})
\frac{1}{\omega^{n+2}}$.  
Another equivalent method  is to express the polarization function
in terms of $\Pi(Q^2, {\bf q}^2)$ and extract the linear ${\bf q}^2$ term.

\section{Linear density Approximation}

In the linear density approximation, 
\begin{eqnarray}
\label{ope5}
\langle G| A | G \rangle =\langle 0| A | 0 \rangle
 + \frac{\rho_n}{2 m} \langle p| A | p\rangle,
\end{eqnarray}
where the first term denotes the vacuum expectation value, which vanishes 
for operators with spin,  and the second
term the nucleon expectation value with the normalization 
 $\langle p|   p\rangle=2p_0 \delta^3(p-p)$. $\rho_n,m$ denotes the 
nuclear density and the mass of the nucleon respectively. 

As in the vacuum, we will truncate our OPE at dimension 6 operators.  This  
implies that in our OPE in eq.(\ref{ope4}), we will have contributions from 
 $(\tau,s)=(2,2),(2,4),(4,2)$.   The nucleon 
matrix elements of the $\tau=2$ operators are very precisely known.  
The $\tau=4$ matrix elements appearing in the $\rho,\omega$ sum rule are 
similar to those appearing in electron DIS\cite{Jaffe} and have been 
estimated  \cite{CHKL93,L94}up to about $\pm$ 30\% uncertainty from 
available DIS data from CERN and Slac.  

The final form of the OPE for  $\Pi^1$ looks as follows.
\begin{eqnarray}
\Pi_{L,T}^1(\omega)/\rho_n= {b_2 \over \omega^6} + {b_3 \over \omega^8}.
\label{ope6}
\end{eqnarray}
 For $\rho, \omega$, the transverse  (T) and longitudinal (L) parts give,  
\begin{eqnarray}
b_2^T & =& (\frac{1}{2}C_{2,2}^q -\frac{1}{2}C_{L,2}^q) mA_2^{u+d}  + 
(C_{2,2}^G -C_{L,2}^G) mA_2^{G}, \nonumber \\ 
b_3^T & = & (\frac{9}{4}C_{2,4}^q -\frac{5}{2}C_{L,4}^q) m^3A_4^{u+d}  + 
(\frac{9}{2}C_{2,4}^G -5C_{L,4}^G) m^3A_4^{G} \nonumber \\
& &  + \frac{1}{2} m \left(
-(1+\beta)(K^1+\frac{3}{8}K^2+\frac{7}{16}K^g)+K^1_{ud}(1\pm1)
\right), \nonumber \\
b_2^L & =&  -\frac{1}{2}C_{L,2}^q mA_2^{u+d}  -C_{L,2}^G mA_2^{G}, \\ 
b_3^L & = & (\frac{1}{2}C_{2,4}^q -\frac{5}{2}C_{L,4}^q) m^3A_4^{u+d}  + 
(C_{2,4}^G -5C_{L,4}^G) m^3A_4^{G} \nonumber \\
& & + \frac{m}{8}(1+\beta) 
\left( K^2-\frac{3}{2}K^g \right),
\label{ope7}
\end{eqnarray} 
where $\pm$ refers to  the $\rho$ and $\omega$ case. 
Here, for even $n$,  
\begin{eqnarray}
A_n^q  = 2 \int_0^1 dx x^{n-1} [ q(x,Q^2)+ \bar{q}(x,Q^2)]~~,  ~~
A_n^G  =  
2 \int_0^1 dx x^{n-1} G(x,Q^2),
\label{ope8}
\end{eqnarray}
where $q(x,Q^2)$ and $G(x,Q^2)$ are the quark and gluon distribution functions.
We will use the HO parameterization for these obtained in ref\cite{Reya92} 
which should be used 
with the Wilson coefficients $C's$ in the ${\overline {\rm MS}}$ 
 scheme\cite{Muta78} and given below.  
\begin{table}[hbt]
% -----------------------------------------------------
% adapted from TeX book, p. 241
\newlength{\digitwidth} \settowidth{\digitwidth}{\rm 0}
\catcode`?=\active \def?{\kern\digitwidth}
% -----------------------------------------------------
\caption{Wilson coefficients for $\tau=2$. 
$T(R)=\frac{f}{2},C_2(R)=\frac{4}{3}$.  
\label{tab:wilson}}
\begin{tabular}{l  l l}
\hline \\
 Wilson coefficient  & $n=2$ & $ n=4$ 
\\
\hline
 $ C_{2,n}^q = 1+ \frac{\alpha_s}{4\pi} B_{2,n}^{NS} $   &
 $ 1+ \frac{\alpha_s}{4\pi}( 0.44) $           & 
 $ 1+ \frac{\alpha_s}{4\pi}(6.07) $  
\\
 $ C_{L,n}^q = \frac{\alpha_s}{4 \pi} C_2(R) \frac{4}{n+1} $  &
 $ \frac{\alpha_s}{4\pi}\frac{16}{9}  $  & 
 $ \frac{\alpha_s}{4\pi}\frac{16}{15} $ 
\\
 $ C_{2,n}^G = \frac{\alpha_s}{4 \pi} T(R) \frac{4}{f} 
  [ \frac{4}{(n+1)} - \frac{4}{(n+2)} + \frac{1}{n^2}
  - \frac{n^2+n+2}{n(n+1)(n+2)}(1+\sum_{j=1}^n \frac{1}{j} )] $  &
 $ \frac{\alpha_s}{4\pi}(-\frac{1}{2})  $ & 
 $ \frac{\alpha_s}{4\pi}(-\frac{133}{180}) $ 
\\
 $ C_{L,n}^G =
  \frac{\alpha_s}{4 \pi} T(R) \frac{1}{f} \frac{16}{(n+1)(n+2)} $  & 
 $ \frac{\alpha_s}{4\pi} \frac{2}{3}  $  & 
 $ \frac{\alpha_s}{4\pi} \frac{4}{15} $ \\
\hline
\end{tabular}
\end{table}
 Terms proportional to $K's$ come from $\tau=4,s=2$.   We will use the set of
 K values obtained in ref.\cite{CHKL93}; 
( $K^1,K^2,K^1_{ud},K^g)=(-0.173{\rm GeV}^2,0.203{\rm GeV}^2,
-0.083{\rm GeV}^2,-0.238{\rm GeV}^2)$ and 
we take $\beta=0.5$.  For the $\tau=4$ operators, we neglect the $Q^2$ 
dependence.

\section{Constraints}
We will now make a Borel transformation of $\omega^2$ times 
eq.(\ref{ope6}) and look at the dispersion relation.
\begin{eqnarray}
\rho_n \left( 
 \frac{b_2}{M^2} -\frac{b_3}{2M^4}  \right)
= \int ds  \rho^1(s)s e^{-s/M^2} +b_{scatt}.
\label{ope9}
\end{eqnarray}
The reason for multiplying by $\omega^2$ is to get rid of any possible 
subtraction constants proportional to $1/\omega^2$.  There could be still 
further subtraction constants proportional to $b_{scatt}/\omega^4$, which 
we included.  $b_{scatt}$ should also be calculated in any model calculations.
 The value from particle-hole intermediate state gives
 $b_{scatt}=-1/(4m) \rho_n$ 
for the longitudinal $\rho,\omega$ meson and zero for the transverse parts.  

After the Borel transformation, the contribution from dimension=$n$ operators
 in the OPE side is  further divided by  $(n-2)!$.  
This will make our truncation at dimension 6 
operators valid even to smaller Borel mass region.  
In fig.1, we have plotted OPE as a function of
 $M^2$.  
  The solid line shows the left hand side of eq.(\ref{ope9}).  
The dot-dashed line
without the twist-4 contribution and the long dashed line without dimension 6
contribution.   
At $M_{min}^2 \sim .8 {\rm GeV}^2$ 
($M_{min}^2 \sim 2.5 {\rm GeV}^2$), 
the contribution of dimension 6 operators become
about 40\% of the contributions from dimension 4 operators for transverse
(longitudinal) polarization.  Therefore, the 
OPE should be reliable only above these minimum values of the Borel
 mass, where the contributions from  higher dimensional operators should be 
similarly suppressed.  To support our argument, we have also plotted (short
dashed line) the 
contributions from adding up the infinite sum of the twist-2 operators.
This is possible because for the twist-2 operators, all the moments are know.
 As can be seen from the figure, once we are above the 
minimum Borel mass, the difference to the solid line is vanishingly small,
confirming the validity of our OPE above the  Minimum Borel mass.
\begin{figure}[htb]
\begin{minipage}[t]{77mm}
\vbox to 2.3in{\vss
   \hbox to 1.3in{\includegraphics{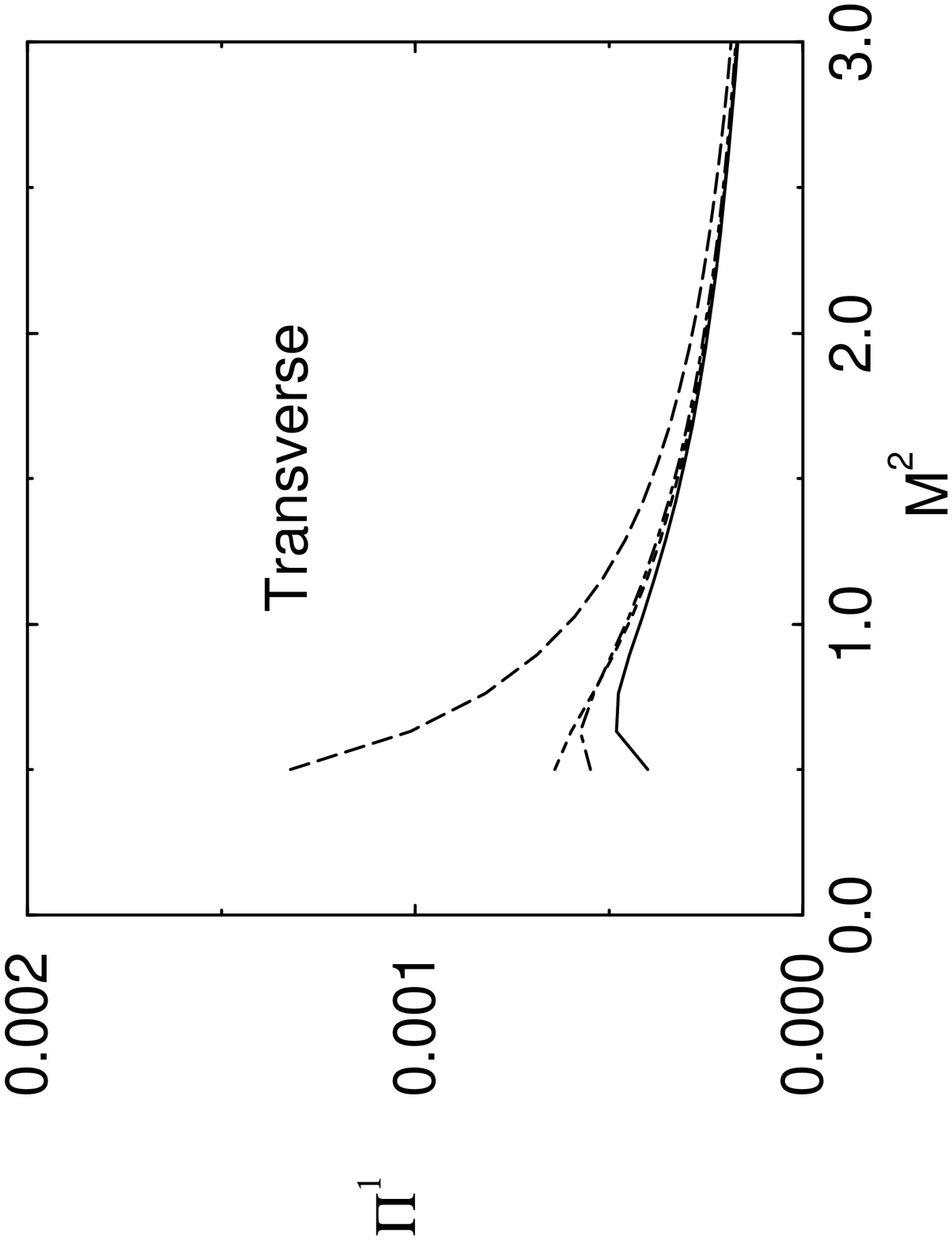}\hss}}
%\caption{Transverse direction.}
\label{fig:largenenough}
\end{minipage}
\hspace{\fill}
\begin{minipage}[t]{77mm}
\vbox to 2.3in{\vss
   \hbox to 1.3in{\includegraphics{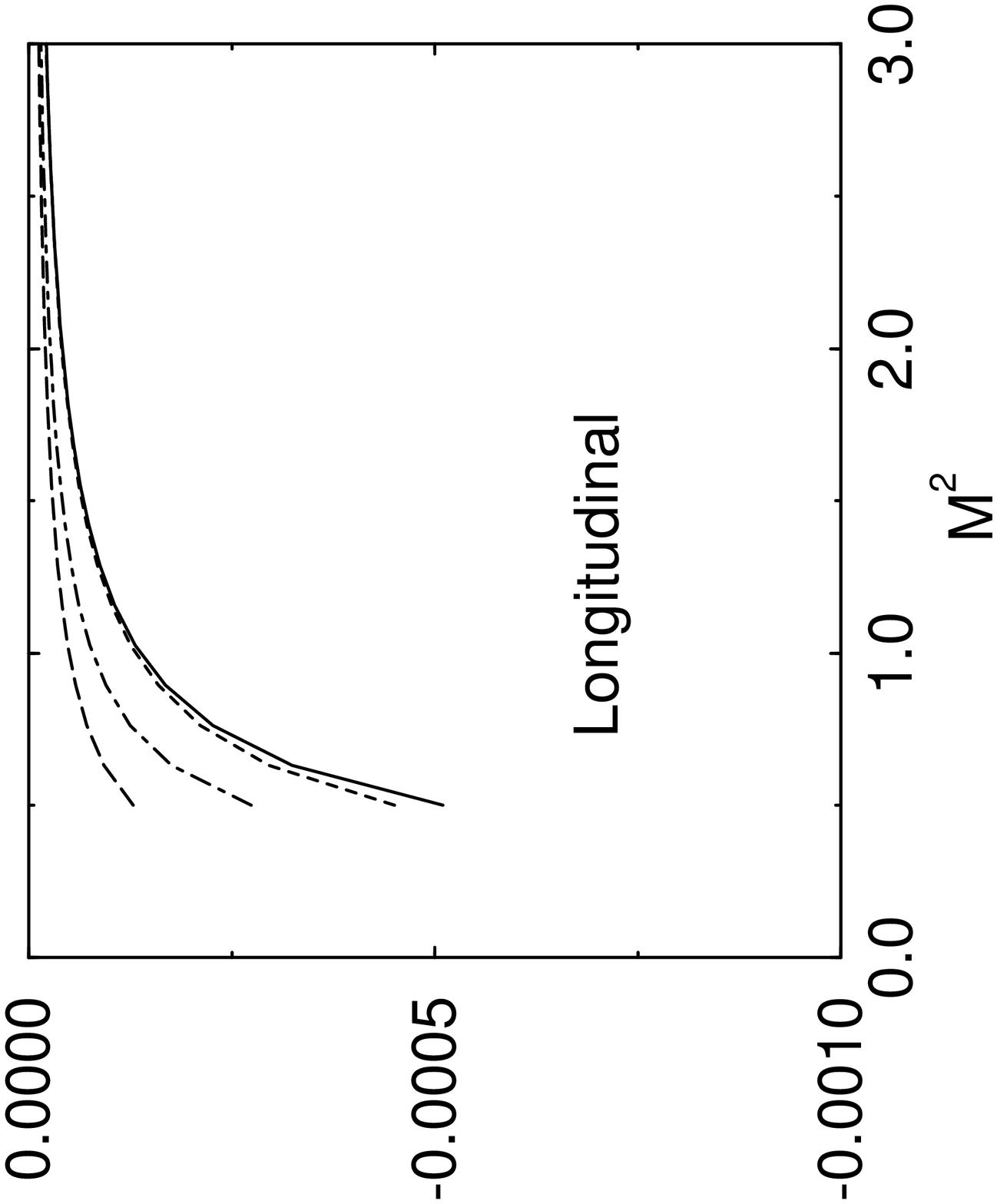}\hss}}
%\caption{Solid line denotes our OPE, dotted line 
%with the infinite sum of the twist-2 operators, dashed line OPE without 
%dim 6, dot-dashed OPE without twist-4}
\label{fig:l1}
\end{minipage}
\caption{Solid line denotes our OPE, short dashed line 
with the infinite sum of the twist-2 operators, dashed line OPE without 
dim 6, dot-dashed OPE without twist-4. All curves are at nuclear matter density
and GeV${}^2$ unit for both axis}
\end{figure}
Now the constraints for the spectral density would be eq.(\ref{ope9}) , but 
applied only above $M^2> M^2_{min}$.  

As can be seen in eq.(\ref{ope9}), the Borel transformation also changes the 
weighting factor of the spectral 
density  to an $exp(-s/M^2)$.   This has the following advantage
for practical applications of our constraint.  For small values of the 
Borel mass, the contribution of the spectral density at larger energy is 
exponentially suppressed.  Consequently, in a model calculation, one can 
concentrate on the change of the spectral density near the vector meson mass
region and below and model the higher energy changes with a simple pole like 
contribution.

\subsection{Results with simple ansatz}

We now show the result with the following ansatz of the spectral density.
\begin{eqnarray}
\rho(u)= { F \Gamma_\rho (m_\rho+ \frac{f}{m_\rho} {\bf q}^2) 
\over (u^2 -m_\rho^2 -a {\bf q}^2)^2  +m_\rho^2 (\Gamma_\rho^2 
+ g {\bf q}^2) } + c \theta(u^2-(S_0+s {\bf q}^2)).
\label{ope10}
\end{eqnarray}
We expand this to linear order in ${\bf q}^2$ to obtain $\rho^1$ and 
use the constraints in eq.(\ref{ope9}) to obtain the best
fit value of the constants $f,a,g,s$.  
In the limit $\Gamma \rightarrow 0$ and $g=0$, the results were obtained in
ref.\cite{L98c}\footnote{The numbers in \cite{L98c} were obtained with incorrect
sea quark distribution.  With the correction, we get slightly larger number
given in this work} which translates into the following momentum dependence 
in the mass. $\frac{m_\rho(\rho_n)}{m_\rho(0)}=1-(0.023 \pm0.007) (\frac{{\bf q}}{0.5})^2
 (\frac{\rho_n}{\rho_0})$, where ${\bf q}$ is in GeV unit and $\rho_0$ is the
 nuclear matter density.  For the $\omega$ meson, 0.023 changes to 0.016.

\section{Conclusion}

We have shown that there are model independent constraints on the 
three momentum dependence of the vector meson spectral density in 
nuclear medium.  Any model calculations can be checked to see if it 
satisfies the constraints.
 Phenomenological models based on p-wave nucleon resonances seems to 
overestimate the momentum dependence\cite{FL98}.

\section{Acknowledgments}

This work was supported in part by the KOSEF through grant no. 971-0204-017-2 and
by the 
Korean Minsitry of Education through grant no. BSRI-98-2425.

\end{document}